\documentclass[12pt,a4paper]{article}
  \usepackage{a4wide}
  \usepackage{latexsym}
  \usepackage{epsf}
  \usepackage{amssymb}
  \usepackage{graphicx}
  \usepackage{amsmath, cite}
 \usepackage{bbm}
  \usepackage{amsmath,amssymb,amsthm}
  \usepackage{verbatim}
  \usepackage{hyperref}
  \usepackage{amsfonts}
\renewcommand{\d}{\textrm{d}}
\newcommand{\e}{\textrm{e}}

\newcommand{\be}{\begin{equation}}
\newcommand{\ee}{\end{equation}}
\newcommand{\ba}{\begin{eqnarray}}
\newcommand{\ea}{\end{eqnarray}}

\newcommand{\sgn}{\textrm{sgn}\,}

\newcommand\varpm{\mathbin{\vcenter{\hbox{%
  \oalign{\hfil$\scriptstyle+$\hfil\cr
          \noalign{\kern-.3ex}
          $\scriptscriptstyle({-})$\cr}%
}}}}

\newcommand\varmp{\mathbin{\vcenter{\hbox{%
   \oalign{\hfil$\scriptstyle-$\hfil\cr
           \noalign{\kern-.3ex}
          $\scriptscriptstyle({+})$\cr}%
}}}}

\newcommand\varbs{\mathbin{\vcenter{\hbox{%
   \oalign{\hfil$\scriptstyle>$\hfil\cr
           \noalign{\kern-.3ex}
          $\scriptscriptstyle({<})$\cr}%
}}}}

\newcommand\varsb{\mathbin{\vcenter{\hbox{%
   \oalign{\hfil$\scriptstyle<$\hfil\cr
           \noalign{\kern-.3ex}
          $\scriptscriptstyle({>})$\cr}%
}}}}

\begin{document}
\numberwithin{equation}{section}
\begin{flushright}
\small IPhT-T12/036\\
\small ITP-UH-09/12\\
\date \\
\normalsize
\end{flushright}
\vspace{0.4cm}
\begin{center}

{\LARGE \bf{Persistent anti-brane singularities}} \\

\vspace{1.6 cm} {\large  Iosif Bena$^\flat$, Daniel Junghans$^\ddagger$, Stanislav Kuperstein$^\flat$ \\
\vspace{0.2cm}Thomas Van Riet$^\flat$, Timm Wrase$^\sharp$ and Marco
Zagermann$^\ddagger$}\footnote{iosif.bena @ cea.fr, daniel.junghans
@ itp.uni-hannover.de, stanislav.kuperstein @ cea.fr,  thomas.van-riet @ cea.fr, timm.wrase @ cornell.edu, marco.zagermann @ itp.uni-hannover.de }\\

\vspace{1.5 cm} {$\flat$ Institut de Physique Th\'eorique, CEA
Saclay, CNRS URA 2306 \\ F-91191 Gif-sur-Yvette, France}

\vspace{.15 cm}  {${}^\ddagger$ Institut f{\"u}r Theoretische Physik \&\\
Center for Quantum Engineering and Spacetime Research\\
Leibniz Universit{\"a}t Hannover, Appelstra{\ss}e 2, 30167
Hannover, Germany}\\

%\vspace{0.2cm} {\upshape\ttfamily daniel.junghans,
%marco.zagermann@itp.uni-hannover.de} \\

\vspace{.15 cm}  {${}^\sharp$ Department of Physics, Cornell University, Ithaca, NY 14853, USA}\\

%\vspace{0.2cm} {\upshape\ttfamily timm.wrase@cornell.edu} \\

%\vspace{0.2cm} {\upshape\ttfamily thomas.van-riet@cea.fr} \\

\vspace{1.2cm}
{\bf Abstract}
\end{center}

\begin{quotation}
\noindent Anti-D-branes inserted in warped throat geometries
(supported by fluxes that carry D-brane charges) develop unphysical
singularities. It has been argued that these singularities
could be resolved when one goes beyond the linearized approximation
or includes the effects of brane polarization. In this
paper we consider anti-D6 branes, whose singularities have been
shown to exist at the full non-linear level, and demonstrate that
there is no D8 brane polarization that can resolve the singularity.
We comment on the potential implications of this result for the
resolution of anti-D3 brane singularities in the Klebanov-Strassler
geometry.
\end{quotation}

\newpage

%\tableofcontents

\section{Introduction}
Anti-branes in warped throat geometries are essential ingredients in
many standard scenarios for moduli stabilization in de Sitter space
\cite{Kachru:2003aw, Balasubramanian:2005zx} or in holographic duals to
non-supersymmetric gauge theories \cite{Kachru:2002gs}. Since an
anti-brane preserves different supercharges than the fluxes and
(orientifold and D-brane) sources that sustain the throat, it feels
a non-zero force that brings it to the bottom of the throat. At the
bottom the tension and charge of the anti-brane is strongly
redshifted and, at least far enough from it, one can hope to view the
anti-brane as a perturbation on top of a supersymmetric background.

In the Klebanov--Strassler (KS) warped deformed conifold solution
\cite{Klebanov:2000hb} one can demonstrate that if the anti-brane backreaction is ignored the
anti-brane feels a classical barrier against annihilation with its surrounding flux,
and hence describes a meta-stable state of the dual theory. The annihilation
proceeds through the polarization \cite{Myers:1999ps} of the anti-D3 branes
into a spherical NS5 brane wrapping a two-sphere of finite radius \cite{Kachru:2002gs} inside the large three-sphere at the bottom of the KS solution. A similar phenomenon has been found for anti-M2 branes \cite{Klebanov:2010qs} in backgrounds with M2 brane charge dissolved in fluxes \cite{Cvetic:2000db}.

It is clearly important to investigate the backreaction of these anti-branes and see whether the existence of a metastable state is a real feature of the system, or an artifact of the probe approximation. There are two methods to find the answer: the first is to study the anti-branes as non-supersymmetric perturbations of the BPS solution; the second is to calculate their full backreaction and see if the resulting solution makes sense.
The first method has been first applied to anti-D3 branes in the KS throat \cite{Bena:2009xk}, and revealed the (surprising) existence of a singularity in the IR of the solution coming from the infinite energy-momentum density in the fluxes. Further investigations of solutions with anti-M2  \cite{Bena:2010gs, Massai:2011vi} and anti-D2 branes \cite{Giecold:2011gw} have shown that such a singularity in the fluxes is ubiquitous. Moreover, unlike the anti-D3 singularity, whose infinite energy-momentum density is integrable, the anti-M2 singularity is not integrable, and the singularity of the fluxes near the anti-D2 branes is even stronger than that of the branes themselves! Hence, it is clear that this singularity cannot be accepted as it is.

The second method to study anti-branes in flux backgrounds is to calculate their full backreaction, which is much more challenging and so far this has only been done for anti-D6 branes \cite{Blaback:2011nz, Blaback:2011pn}. This investigation also revealed the presence of a singularity coming from the infinite energy density of the fluxes near the anti-branes. Upon T-dualizing the anti-D6 brane solution three times along the anti-D6 brane worldvolume one
obtains an anti-D3 brane solution in the non-compact Calabi-Yau
$\mathbb{T}^3 \times \mathbb{R}^3$, where the anti-D3 branes are smeared
over the 3-torus $\mathbb{T}^3$ and the effect of T-duality is to
map the Romans mass $F_0$ into $F_3$-flux filling the
$\mathbb{T}^3$. This way one obtains a singular anti-D3 brane solution,
whose singularities are of the same type as those of anti-D3 branes in a KS throat \cite{Massai:2012jn}.
This is a strong indication that the anti-brane singularities found in perturbation theory remain when one considers the full backreaction, contrary to claims by \cite{Dymarsky:2011pm}.

Given the likely persistence of anti-brane singularities at the full nonlinear level, one can ask what do these singularities indicate. One possibility is these singularities are unphysical, and hence there is no solution for anti-branes in KS \cite{Bena:2009xk}. They can indicate for example the existence of a perturbative decay channel for the anti-branes against the surrounding flux\cite{Blaback:2012nf}.

The other possibility is that these singularities will be resolved if one considers the full non-perturbative dynamics of the anti-branes, and in particular their polarization into NS5 or D5 branes, the way it happens in the Polchinski-Strassler (PS) solution \cite{Polchinski:2000uf}. So far the evidence for singularity resolution by brane polarization is split: on one hand, in the probe approximation the anti-D3 branes do polarize into an NS5 brane \cite{Kachru:2002gs}. On the other hand, the fluxes that give rise to the singularity are not of the correct type and orientation to couple to such NS5 brane.

Unfortunately ascertaining whether the singularity of the anti-brane solutions constructed so far is resolved by brane polarization is not easy. For example the anti-D3 brane perturbative solution \cite{Bena:2009xk, Bena:2011wh, Bena:2011hz} \footnote{See \cite{Kuperstein:2003yt, McGuirk:2009xx, DeWolfe:2008zy} for earlier work.} is a cohomogeneity-one solution constructed using the Borokhov-Gubser method \cite{Borokhov:2002fm}, and describes anti-D3 branes that are smeared over the $S^3$ of the conifold. This smearing wipes out the NS5 polarization channel found in \cite{Kachru:2002gs}. Similarly, the smearing of the anti-M2 over the tip of CGLT background \cite{Cvetic:2000db} wipes out the M5 polarization channel of \cite{Klebanov:2010qs}.

To make progress one has to remember an old piece of brane polarization lore: in PS a D3 brane that polarizes always has {\it two} polarization channels, corresponding to D5 and NS5 polarization in orthogonal planes\footnote{A similar situation exists in other brane polarizations that can be studied {\it \`a la PS} or by constructing the fully backreacted solution, like M2 branes polarizing into M5 branes in orthogonal planes \cite{Bena:2000zb, Bena:2004jw, Lin:2004nb} or D2 branes polarizing into D4 and NS5 branes in orthogonal planes \cite{Bena:2000fz}.}. When the branes that are polarizing are fully backreacted, both these channels are visible. On the other hand, if one considers the branes that are polarizing as probes in a certain flux background, only one channel is captured by the probe approximation, while the other one is generically not.

If one applies this fact to antibranes in KS, one expects that besides the NS5 brane polarization channel visible when the anti-D3 branes are considered in the probe approximation \cite{Kachru:2002gs}, there should exist another polarization channel in which the anti-D3 branes polarize into a D5 brane wrapping the $S^2$ of the deformed conifold and sitting at a finite distance from the tip. However, this D5 channel is not visible when one considers the anti-D3 branes as probes and ignores their backreaction, because the RR vector potential $C_6$ vanishes in the unperturbed KS background.
% Even if one goes one step beyond the probe approximation, and considers the first-order
%  backreaction of the anti-branes, this D5 channel is still not visible.
Hence, the only way to see whether the D5 polarization channel exists is to consider solutions where the anti-D3 branes are fully backreacted. Moreover, since the polarization happens in a plane orthogonal to the smearing direction, this channel - if existent - will survive the smearing.

Unfortunately, the fully backreacted smeared anti-D3 brane solution in KS has not been found yet, and hence
performing the full analysis is not possible. Nevertheless, what is known is the solution for anti-D3 branes on $\mathbb{T}^3 \times \mathbb{R}^3$, which is T-dual to the anti-D6 solution of \cite{Blaback:2011nz, Blaback:2011pn}. This solution captures the essential physics of the region near the anti-D3 branes in KS when the ``Schwarzschild radius" of these anti-branes ($\sim g_s \overline{N}_3$) is much much smaller than the size of the $S^3$. Hence, if one could prove or disprove the existence of a D5 brane polarization channel for anti-branes in $\mathbb{T}^3 \times \mathbb{R}^3$, it would be very surprising if the physics of this channel in KS would be any different. On the other hand, since the solution of \cite{Blaback:2011nz, Blaback:2011pn} depends on several parameters that can only be fixed from the UV physics, so it is also possible a-priori that polarization happens when these parameters are in a certain range, and does not happen otherwise.

The result we find is quite surprising: in the solution of \cite{Blaback:2011nz, Blaback:2011pn}  the polarization potential has exactly the same terms that appear when branes polarize intro branes two dimensions higher \cite{Polchinski:2000uf, Bena:2000fz,Bena:2001aw}, but the coefficients of these terms are such that there is no polarization! \footnote{Another example where there is no polarization despite the presence of all the terms in the polarization potential is the Pilch-Warner solution \cite{Pilch:2000ue}. This can be seen by taking one of the chiral multiplet masses to zero in \cite{Polchinski:2000uf}: in this limit the shape of the D5 and NS5 branes in which the D3 branes polarize degenerates, and the D5 and NS5 dipole charges disappear.}

Furthermore, this result is independent of the free parameters of this solution, which indicates that the absence of polarization is universal. The obvious conclusion of this calculation is that brane polarization does not seem to help in curing the infrared singularities of anti-branes in backgrounds with charge dissolved in fluxes, and hence this singularity is there to stay.

Since the T-dual of our geometry can be argued to capture the near-tip physics of anti-D3 branes in KS, our calculation also indicates that the D5 brane polarization channel will be absent for smeared anti-D3 branes. Since this polarization channel is in a plane transverse to the smearing direction, it will likely also be absent for localized anti-D3 branes, which in turn would indicate that its ``twin'' NS5 channel is also absent. Thus it seems that brane polarization will probably not help cure the singularities of anti-D3 branes in warped throats with fluxes. It would be very important to confirm this calculation on a fully-backreacted anti-D3 KS solution.

This paper is organized as follows: In Section \ref{prel} we
establish our notation by recalling the essential properties of the
anti-D6 brane solution described in \cite{Blaback:2011nz, Blaback:2011pn}.
In Section \ref{sec:D8potential} we compute the potential for the D6 branes to polarize into a D8 brane wrapping an $S^2$, or equivalently of anti-D3 branes on $\mathbb{T}^3 \times \mathbb{R}^3$ to polarize into a D5 brane, and demonstrate the absence of a meta-stable
state.  In the appendix we present the details of the analogous calculation
for the compact setting.

\section{Preliminaries} \label{prel}

\subsection{The Ansatz}

In order to obtain a solution with D6 brane charge dissolved in fluxes, the space transverse to the branes must have a nonzero NS-NS three-form field strength, $H_3$, as well as a nonzero Romans mass, $F_0$. If one considers a solution with D6 or anti-D6 branes located at a point in the transverse space, the solution has SO(3) symmetry, and is described by the string-frame Ansatz \cite{Blaback:2010sj}:
\begin{eqnarray}
\label{TheBackground}
\d s^2_{10} & = &
   \e^{2 A(r)+\tfrac{1}{2}\phi(r)}\d x_\mu \d x^\mu +\e^{2B(r)+\tfrac{1}{2}\phi(r)}\Big( \d r^2 + r^2 (\d \psi^2 + \sin^2 \!\psi\, \d \phi^2 )\Big)\, , \\
H_3 & = & F_0 \lambda(r)  \e^{\tfrac{7}{4}\phi(r) + 3B(r)} r^2 \d r \wedge \omega_{S^2} \, ,  \label{TBg2} \\
F_2 & = & \e^{-\tfrac{3}{2}\phi(r) - 7 A(r) + B(r)} \alpha'(r) r^2 \,\omega_{S^2} \, , \label{TBg3}\\
F_0 & \neq & 0,\label{TBg4}
\end{eqnarray}
where  $F_0$ is the value of the Romans mass, and the equations of motion governing the five unknown
functions $A(r), B(r), \phi(r), \alpha(r)$ and $\lambda(r)$ are ODE's.
The D6 worldvolume coordinates are
$x^{\mu}$ ($\mu=0 \ldots 6$), $\omega_{S^2} = d \psi \wedge \sin \!\psi\, \d \phi$ is the volume form of the transverse two-sphere, and
the $H_3$-flux is proportional to the volume element of the three-dimensional
transverse space.
% (the choice to write its  ($r$-dependent) factor of
%proportionality as $\lambda F_0 \e^{\tfrac74\phi}$ instead of $\lambda$,
%turns out more useful. The same applies to the expression of $F_2$.
%that could equally well have been given by $F_2 =\star_3\d\alpha$.

The extremal\footnote{The solution is only supersymmetric when the worldvolume symmetries are broken by a adding a term linear in a spatial worldvolume coordinate to the harmonic function.} D6 brane background \cite{Janssen:1999sa} has:
\begin{equation}
\lambda=1\,,
\end{equation}
which is the correspondent of the familiar D3-brane ISD-flux condition in general D-brane backgrounds, and can be related to
the latter by T-duality. The metric, dilaton and two-form field strength are exactly the same as for supersymmetric D6 branes in flat space,
\begin{equation}
\e^{2A} =h_6^{-\frac{1}{8}}\,,\qquad \e^{2B} = h_6^{\frac{7}{8}}\,,\qquad \e^{\phi} = g_s h_6^{-\frac{3}{4}}\,,\qquad \alpha =\e^{\tfrac{3}{4}\phi +7A}\,,\end{equation} except that the D6 harmonic function now picks up a contribution from the D6 charge dissolved in the fluxes:
\begin{equation}
 h_6(r) = 1 + \frac{g_s \tilde{N}_6 }{r} -
\frac{1}{6} F_0^2 r^2 \, . \label{D6warpfunction}
\end{equation}
where $\tilde{N}_6 \equiv  \frac{\pi}{2} N_6$, and we have set  $l_s=1$.

The corresponding warp factor matches exactly the infrared behavior of the Klebanov-Strassler warp factor with mobile branes smeared on the KS three-sphere, which confirms our intuition that it captures the IR behavior of the KS near-tip region.
Far far away from the origin the warp factor vanishes, and the solution has a naked singularity, but this does not  need to bother us. If one is interested in this solution for its ability to capture the essential IR physics of anti-branes in KS, we know that in KS there will be other contributions to the warp factor when one goes further away from the tip, and hence this singularity will not be there. If we are interested in a D6 solution by itself, the problem can be cured in a compact setting, as it provides a natural cut-off. A
compact solution can be obtained by changing the D6 branes into an O6 plane, which
is a solution T-dual to the well-known GKP solution for O3 planes
\cite{Giddings:2001yu}\footnote{A more involved procedure is to glue
a compact space, containing orientifolds, to the D6 brane throat.}.

We cannot introduce anti-D6 branes directly on top of this background since
they would annihilate against the D6 brane source.  This is due to the fact that the background does not allow a resolution of the brane singularity. However, if no D6 branes are present, one can insert an anti-D6 brane
in a flux background that carries D6 brane charge. This implies that far enough from
the anti-D6 brane source one demands that the fluxes go to their BPS value
for D6 branes, or equivalently, far enough (but still before
one hits the even further away singularity) we should have:
\begin{equation}\label{UV}
\lambda_{\text{UV}} \rightarrow 1\,.
\end{equation}
This is the only ``UV'' boundary condition we impose.

\subsection{The singularity and its possible resolution}

Before constructing the full-blast anti-D6 brane solution, one can already argue that this solution will have an infrared singularity, coming from the divergence of the energy density of $H_3$ at $r=0$ \cite{Blaback:2011nz}.
%, similar to the one found in the first-order backreaction calculations of \cite{Bena:2009xk, Bena:2011hz}.
This argument proceeds in three steps:
\begin{enumerate}
  \item The $H_3$ equation of motion gives an algebraic
    relation between $\alpha$ and $\lambda$:
    \begin{equation}\label{alphaversuslambda}
        \alpha=\e^{\tfrac{3}{4}\phi +7A}\lambda +\alpha_0\,.
    \end{equation}
    where $\alpha_0$ is a gauge choice, which we take zero for now.
    Hence, our UV (large $r$) boundary condition (\ref{UV}) implies that:
    \begin{equation}\label{UV2}
        \alpha_{\textrm{UV}} > 0 \,.
    \end{equation}
  \item In the IR we must have:
    \begin{equation}\label{derivativeIR}
        \alpha^\prime_{\textrm{IR}} < 0\,,
    \end{equation}
    since the derivative of $\alpha$ (denoted by $\alpha^\prime$) near the
    source is determined by the D6 brane charge, which in our conventions is negative for an anti-D6 brane.
  \item From a combination of the $F_2$ Bianchi identity and (\ref{alphaversuslambda}) one can demonstrate
    that if $\alpha^\prime (r_0) = 0$ for some $r=r_0$ then necessarily:
    \begin{equation}\label{extremum}
    \sgn \left( \alpha(r_0) \right) = \sgn \left( \alpha^{\prime \prime}(r_0) \right) \,.
    \end{equation}
\end{enumerate}
By combining (\ref{UV2},  \ref{derivativeIR}, \ref{extremum})
one can fix the overall behavior (``topological'' properties) of the
function $\alpha(r)$ . The result is that $\alpha(r)$ necessarily reaches
a strictly positive (non-zero) value at $r=0$ \cite{Blaback:2011nz}:
\begin{equation}
\alpha_{\textrm{IR}} >0 \,.
\end{equation}
This is enough to demonstrate the singularity in the H-flux energy density
\begin{equation}
\e^{-2\phi}H^2 = \alpha^2\e^{-\tfrac32\phi-14A}F_0^2\,.
\end{equation}
If we evaluate this quantity in the IR, where the dilaton and the
warp factor behave as for ordinary (anti-)D6 branes in flat
space\footnote{This assumption has been explicitly verified in
\cite{Blaback:2011pn}, where the existence of the singularity in the H-flux density was established in Einstein frame.} we find that
\begin{equation}
\label{IRbehavior}
\e^{\phi_{\textrm{IR}}} \sim r^{3/4}\,,\qquad \e^{A_{\textrm{IR}}} \sim r^{1/16}\,,
\end{equation}
and hence near $r=0$
\begin{equation}
\e^{-2\phi} H^2 \sim \alpha_{\textrm{IR}}^2 r^{-2}\,.
\end{equation}
This clearly diverges since $\alpha_{\textrm{IR}} \neq 0$.

Let us now show how the singularity could be cured if the anti-D6 branes polarize into D8 branes.
When we replace the IR boundary conditions with the boundary conditions near a D8 brane shell wrapping an $S^2$
at finite radius ($r=r^*$), we still find that $\alpha(r^{*})>0$, but near the D8 brane the warp factor does not blow up, and hence both the dilaton and the warp factor are finite and
\begin{equation}
\e^{-2\phi} H^2 \sim \alpha(r^*)^2\, ,
\end{equation}
which is everywhere finite.

\subsection{Details of the IR solution}

In this subsection, we analyze in more detail the IR (small $r$)
behavior of the anti-D6 brane background, contained in the Ansatz (\ref{TheBackground})-(\ref{TBg4}). It
was argued in \cite{Blaback:2011pn}  that, apart from the $B_2$-field,
the leading order behavior of all the fields (the metric, the
dilaton and the $F_2$-form flux), near the D6 brane source,  is that
of the BPS D6 brane solution (see the discussion around
(\ref{IRbehavior})). Although the full solution is still out if
reach, it has been demonstrated in \cite{Blaback:2011pn} that one
can consistently solve the equations of motion order by order in the
$r$-expansion. The leading terms in this expansion are:
\begin{eqnarray}
\label{expansion}
e^{-A(r)} &=& r^{-1/16} \left( a_0 + a_1 r + a_2 r^2 + \ldots \right)   \nonumber \\
e^{-2 B(r)} &=& r^{7/8} \left( b_0 + b_1 r + b_2 r^2 + \ldots \right)  \nonumber \\
e^{-\frac{1}{4} \phi (r)} &=& r^{-3/16} \left( f_0 + f_1 r + f_2 r^2 + \ldots \right)  \nonumber \\
\lambda (r) &=& r^{-1} \left( \lambda_0  + \lambda_1 r + \lambda_2 r^2 + \ldots \right)  \, ,
\end{eqnarray}
All (higher order) coefficients depend only on the Romans mass and the five free parameters:
$a_0$, $b_0$, $f_0$,$\lambda_0$, and $\lambda_1$
\cite{Blaback:2011pn}. As we will show below, the first three of these
can be expressed in terms of $g_s$ and $N_6$ (the number of the
anti-D6 branes), while the remaining two parameters, $\lambda_0$,
and $\lambda_1$, cannot be fixed from the IR data only. If one thinks about the T-dual of this solution as
the infrared of the solution describing anti-D3 branes in KS, these coefficients represent information about the
gluing of this infrared to KS, and can only be fixed from the full solution. If one thinks about the compact D6  solution,
these coefficients are again determined by the UV data. While one could have expected that the exact form of the D6
polarization potential depends crucially on these coefficients, this turns out not to happen, and hence whether a
D6 polarizes or not is independent of the UV data !

We can compare the leading terms in (\ref{expansion}) to the dilaton and the warp function of the ``near-horizon" D6 brane BPS
solution. For small $r$, we only need to keep the second, $1/r$, term in the warp function (\ref{D6warpfunction})
\begin{equation}
    h_6(r) \approx \frac{g_s \tilde{N}_6}{r} \, .\label{NH1}
\end{equation}
Requiring that near the source
\begin{eqnarray}
e^{\tfrac{1}{2} \phi + 2 A} \approx h_6^{-1/2}  \qquad
e^{\tfrac{1}{2} \phi + 2 B} \approx h_6^{1/2}   \quad \textrm{and} \quad
e^{\phi} \approx g_s h_6^{-3/4}  \, ,\label{NH2}
\end{eqnarray}
we find that: \footnote{One can check that there is a rescaling of $r$ and the other parameters
of the setup that leaves the gauge fixed equations of motion of
\cite{Blaback:2011pn} invariant such that the expansion (\ref{expansion}) can be
used and consistently matched with the expressions (\ref{NH1}) and (\ref{NH2}).}
\begin{equation}
\label{a0b0f0}
a_0 = g_s^{5/16} \tilde{N}_6^{1/16} \, , \quad
b_0 = g_s^{-3/8} \tilde{N}_6^{-7/8} \, , \quad
f_0 = g_s^{-1/16} \tilde{N}_6^{3/16} \, .
\end{equation}
%It is worth noticing here that we could have used a re-scaling of the radial coordinate
%to arrive at a different result, but we preferred to stick with this gauge choice.\footnote{The re-scaling acts like:
%\begin{equation}
%    r \to \gamma r \, , \quad
%    a_0 \to \gamma^{1/16} a_0 \, , \quad
%    b_0 \to \gamma^{9/8} b_0 \, , \quad
%    f_0 \to \gamma^{3/16} f_0 \, .
%\end{equation}
%}.

\section{The D8 brane potential} \label{sec:D8potential}

In this section we will calculate the potential for the
polarization of the anti-D6 branes into D8 branes by studying the D8
brane worldvolume theory.
More specifically we will compute the action for a probe D8 brane with $n$ anti-D6 branes in the infrared of a solution sourced by
$N_6$ anti-D6 branes.

 As mentioned in the Introduction, this polarization channel is dual to the
D5 polarization channel of anti-D3's in KS, and if this channel is not functional, it is very likely that anti-brane
singularities will not be in general resolved by brane polarization.

To find whether the anti-D6 branes polarize into a D8 brane, we consider a D8 brane that has a large anti-D6 charge dissolved in it, wrapping the topologically trivial $2$-sphere of radius $r=r_\star$ in the $\mathbb{R}^3$ orthogonal to the anti-D6 branes.
A D8 brane with no anti-D6 charge feels a very strong attractive potential, and only has a minimum at   $r_\star=0$.
Since the masses of the D8 and the anti-D6 branes add in quadratures, adding anti-D6 branes makes the
polarization more likely: the anti-D6 DBI attraction and WZ repulsion cancel, and the leftover contribution to the attractive potential coming from the D8 brane is much smaller.

The D8 brane action in string frame is
\begin{equation}
S_{D8} = S_{DBI} + S_{WZ} \,
\end{equation}
with
\begin{equation}
S_{DBI} = -\mu_8 \int \d^9 \xi
e^{-\phi} \sqrt{- \textrm{det} \left( g + 2 \pi \mathcal{F}_2 \right)}  \, ,
\quad
S_{WZ}  = \mu_8 \int \left( C_9 + 2 \pi \mathcal{F}_2 \wedge C_7 \right) \, ,
\end{equation}
where $2 \pi \mathcal{F}_2 \equiv 2 \pi \mathfrak{f}_2 - B_2$ with $\mathfrak{f}_2$ being the usual world volume gauge field strength.

To calculate the potential, we will consider a static configuration with the
D8 brane sitting at a fixed $r$ and spanning the rest of the coordinates.
The induced metric is then:
\begin{equation}
 \sqrt{g_{\parallel}} = e^{7 A + \frac{7}{4} \phi} \cdot \textrm{Vol}_{\mathbb{R}^{1,6}} \, , \qquad
 \sqrt{g_{\perp}} = e^{2 B + \frac{1}{2} \phi} r^2 \cdot \textrm{Vol}_{S^2} \, .
\end{equation}
Furthermore, to give the D8 brane $n$ units of the D6 charge, its world volume field strength has to be
\begin{equation}
\label{f2}
   \mathfrak{f}_2 = \frac{n}{2} \omega_{S^2} \, .
\end{equation}
Using  (\ref{TBg2}),
(\ref{alphaversuslambda}) and the Bianchi identity for $F_2$,  we can write down the $B_2$-form and the RR
vector potentials $C_7$ and $C_9$ (all other forms vanish
identically): \footnote{In our conventions $F_{8} \equiv - \star_{10}
F_{2}$, $ F_{10} \equiv  \star_{10} F_{0}$ and $\d C_{p+2} = F_{p+3}
+ H_3 \wedge C_p$ for $p=1,3,5,7$.}
\begin{eqnarray}
\label{B2C7C9}
  B_{2} &=&  \frac{1}{F_0}  e^{-\frac{3}{2} \phi(r) - 7 A(r) + B(r)} r^2 \alpha^\prime(r)  \cdot  \omega_{S^2} \, , \nonumber \\
  C_7 &=& ( \alpha(r) - \alpha(0) ) \cdot \omega_{\mathbb{R}^{1,6}}  \, , \nonumber  \\
  C_9 &=&  \gamma(r) \cdot \omega_{\mathbb{R}^{1,6}} \wedge \omega_{S^2} \, ,
\end{eqnarray}
where
\begin{equation}
\gamma^\prime(r) = F_0 \left( \alpha(r) ( \alpha(r) - \alpha(0) ) e^{ \phi(r) - 7 A(r) + 3 B(r)} -
                                                    e^{ \frac{5}{2} \phi(r) + 7 A(r) + 3 B(r)} \right) r^2 \, .
\end{equation}
%Here we used (\ref{alphaversuslambda}) in order to find $B_2$ from (\ref{TBg2}).
We also added a constant to $C_7$ so that at large $n$ the
leading contribution at small $r$ of the DBI part will match exactly the one coming from the WZ term (see below).

As explained above, to have any hope for the D8 shell to be stable at a finite radius, the
contribution to the energy coming from the $n$ anti-D6 branes dissolved in it must dominate
over the D8 contribution, which is proportional to the induced metric
$g_\perp$ contribution. This happens in all known examples of brane polarization
(see for example \cite{Polchinski:2000uf, Bena:2000zb, Bena:2000fz}). One can then expand the square
root of the D8 DBI action, and keep the leading terms, proportional to $n$ and $1/n$ respectively.
In a supersymmetric background the term proportional to $n$ cancels the $C_7$ WZ term in the action since
a probe D6 brane should feel no force from a fellow supersymmetric D6 brane. In our solution supersymmetry is broken
and hence the cancelation will be only partial leaving a contribution of order $n r^2$. On the other hand, the $1/n$ term in the DBI expansion
will go like $r^4$. These two terms are necessarily positive, and give an attractive force. On the other hand, the WZ action has also a repulsive $n$-independent term that goes like $r^3$ and comes from the IR expansion of $C_9$. There is no need to keep other terms in the expansion, provided we keep $n$ large and stay at small $r$. Hence, the full polarization potential is
\begin{equation}
\label{D8potential}
    V(r) \sim \left(   \pi n \cdot c_2 r^2 - c_3 r^3 + \frac{1}{  \pi n} c_4 r^4 \right) \, .
\end{equation}
 Note that this is the universal potential for polarization of branes into other branes that are two dimensions higher \cite{Polchinski:2000uf, Bena:2000fz, Bena:2001aw}, and the fact that all the terms are nonzero implies that the fields necessary for polarization are all there\footnote{The sign of $c_3$ can easily be changed by flipping the D8 orientation, and we have chosen the orientation for which the $r^3$ term is negative. For the other orientation all three terms are attractive.}. Hence, whether or not polarization happens depends on the balance between $c_2, c_3$ and $c_4$. For example, if
\begin{equation}
\label{MetaStable}
{32 \over 9} c_2 c_4 < c_3^2 < 4 c_2 c_4 \, ,
\end{equation}
the potential would have a minimum away from the origin, but this vacuum with polarized branes would have higher energy than that of the origin, and would hence be metastable. If however
\begin{equation}
\label{NoPuffingUp}
c_3^2 < {32 \over 9} c_2 c_4 \, ,
\end{equation}
then there will be no minimum away from the origin and the branes will not polarize.

A straightforward calculation yields the following results for the coefficients:
\begin{eqnarray}
c_2 &=& \frac{\lambda_0}{f_0^3 a_0^7} \left( \frac{\lambda_2}{\lambda_0} - 3 \frac{f_2}{f_0} - 7 \frac{a_2}{a_0} -
                       \left( \frac{\lambda_1 + 1}{\lambda_0} \right)^2 + 21 \frac{a_1 f_1}{a_0 f_0} + 6 \left( \frac{f_1}{f_0} \right)^2 +
                       28 \left( \frac{a_1}{a_0} \right)^2 \right) \, ,\nonumber  \\
c_3 &=& \frac{1}{3}  \frac{F_0 \lambda_0}{a_0^7 f_0^{10} b_0^{3/2}} \, , \nonumber  \\
c_4 &=& \frac{1}{2} \frac{1}{(a_0 f_0)^{7} b_0^{2}} \, .
\end{eqnarray}
and it looks at first glance that the polarization potential will depend on many parameters of the solutions, most of which cannot be fixed in the infrared, and will hence depend on the way this solution is embedded in a UV-complete solution. However, a surprise awaits.

We first express all the coefficients in terms of $a_0$, $b_0$, $f_0$, $\lambda_0$, $\lambda_1$ and $F_0$.
For $a_1$, $b_1$ and $f_1$ we have:
\begin{eqnarray}
\label{a1b1f1}
  \frac{a_1}{a_0} &=&  \frac{1}{16} \frac{\lambda_1 + 1}{\lambda_0} - \frac{3}{64} \frac{(F_0 \lambda_0)^2}{b_0 f_0^{10}}
  \nonumber \\
  \frac{b_1}{b_0} &=&  - \frac{7}{8} \frac{\lambda_1 + 1}{\lambda_0} + \frac{31}{96} \frac{(F_0 \lambda_0)^2}{b_0 f_0^{10}}
  \nonumber \\
  \frac{f_1}{f_0} &=&  \frac{3}{16} \frac{\lambda_1 + 1}{\lambda_0} + \frac{7}{64} \frac{(F_0 \lambda_0)^2}{b_0 f_0^{10}} \, .
\end{eqnarray}
It is important to emphasize here that in \cite{Blaback:2011pn} these equations have one extra term on the right hand side, which drops out
when the transverse space to the D6 branes is non-compact. The full equations can be found in the appendix.

Apart from (\ref{a1b1f1}) we will need
one extra relation for the second-order coefficients, which follows from
(B.42), (B.43) and (B.45) of \cite{Blaback:2011pn}
(for some reason it seems that we do not have to use all of the type IIA equations of motion from \cite{Blaback:2011pn}):
\begin{equation}
\label{lambda2a2f2}
    \frac{\lambda_2}{\lambda_0} - 7 \frac{a_2}{a_0} - 3 \frac{f_2}{f_0} = -\frac14 \frac{b_1}{b_0 \lambda_0} +
         \frac{7}{2} \frac{a_1 \lambda_1}{a_0 \lambda_0} - \frac{7}{2} \left( \frac{a_1}{a_0} \right)^2 + 3 \left( \frac{f_1}{f_0} \right)^2
                            + \frac{21}{2} \frac{a_1 f_1}{a_0 f_0} \, .
\end{equation}
From (\ref{a1b1f1}) and (\ref{lambda2a2f2}) we obtain:
\begin{equation} \label{coefficients}
c_2 = \frac{1}{12} \cdot \frac{(F_0 \lambda_0)^2}{a_0^7 b_0 f_0^{13}} \, ,\qquad
c_3 = \frac{1}{3} \cdot \frac{F_0 \lambda_0}{a_0^7 b_0^{3/2} f_0^{10}} \, , \qquad
c_4 = \frac{1}{2} \cdot \frac{1}{a_0^7 b_0^{2} f_0^{7}} \, ,
\end{equation}
and hence Eq. (\ref{NoPuffingUp}) is satisfied and there is no polarization!
%\begin{equation}\label{Delta}
%{32 \over 9} c_2 c_4 > c_3^2\,
% = \frac{1}{18} \frac{(F_0 \lambda_0)^2}{a_0^{14} b_0^3 f_0^{20}} \, .
%\end{equation}

This result is quite remarkable, because $\lambda_1$ drops out from the final expressions of the coefficients, and hence the condition (\ref{NoPuffingUp}) is fulfilled for \emph{any} value of the remaining parameters. In particular, if on thinks about the T-dual of this geometry the representative of the near-anti-brane region at the tip of KS, this calculation indicates that the anti-branes will not polarize irrespective of the values of the parameters of the near-tip solution that depend on UV data. Hence, one cannot make the branes polarize by changing any of the parameters of the solution. The absence of polarization is universal!

\subsection{Regimes of validity}

Let us now return to the discussion of the regime of validity of our calculation.
It is clear from (\ref{D8potential})
that the radius at which polarization would have been possible goes like
\begin{equation}
\label{r-star}
    r_\star \sim \pi n \sqrt{ \frac{c_2}{c_4} } \sim n \cdot F_0 \lambda_0 \frac{b_0^{1/2}}{f_0^3}   \, ,
\end{equation}
and at this radius the three terms in the potential (\ref{D8potential}) are of the same order.

\begin{itemize}
   \item To stay in the probe approximation we take:
    \begin{equation}
    \label{Cond4}
        n \ll N_6 \, .
    \end{equation}
  \item In order to trust the IR series expansion of the solution at the scale at which polarization might take place $r_\star$ has to be smaller than $a_0/a_1$, $b_0/b_1$, $f_0/f_1$ and $\lambda_0/\lambda_1$. Since we do not have the full UV-complete solution, we cannot estimate precisely how the coefficients
    $a_1$, $b_1$, $f_1$, $\lambda_0$ and $\lambda_1$ depend on the free parameters
    $F_0$, $g_s$ and $N_6$.
    Fortunately,  we can still proceed even with no UV completion at hand.
    It is reasonable to assume that for the full solution all three terms on the right hand side of
    (\ref{a1b1f1}) will be of the same order of magnitude. This means that:
    \begin{equation}
        \lambda_1 \sim 1  \quad \textrm{and} \quad
        \frac{a_1}{a_0}, \frac{b_1}{b_0}, \frac{f_1}{f_0} \sim \frac{1}{\lambda_0} \sim \frac{(F_0 \lambda_0)^2}{b_0 f_0^{10}} \, .
    \end{equation}
    Using (\ref{a0b0f0}) we arrive at:
    \begin{equation}
        \lambda_0^3 \sim \frac{N_6}{g_s F_0^2} \, , \quad
        \frac{a_0}{a_1}, \frac{b_0}{b_1}, \frac{f_0}{f_1}  \sim \left( \frac{N_6}{g_s F_0^2} \right)^{1/3} \,
        \quad \textrm{and} \quad
        r_\star \sim n \left( \frac{F_0}{g_s N_6^2} \right)^{1/3} \, . \label{lll}
    \end{equation}
    To conclude, in order to trust the IR series expansion of the solution we require that:
    \begin{equation}
    \label{Cond1}
        n \ll \frac{N_6}{F_0} \, .
    \end{equation}
  \item The DBI square root has been expanded: $\textrm{det} ( 2 \pi \mathcal{F}_2 ) \gg \textrm{det} ( g_\perp )$. This leads to:
    \begin{equation} \label{Cond-DBI}
        n \gg \frac{r_\star^{3/2}}{f_0^2 b_0}  \, .
    \end{equation}
    Plugging in (\ref{lll}) and (\ref{a0b0f0}), we see that somewhat surprisingly this condition reduces to the previous one (\ref{Cond1}).
  \item  The radius of the 2-sphere wrapped by the D8 brane should be large in string units.
    This radius is given by $\sqrt{\textrm{det} ( g_\perp )}$ at $r=r_\star$ and so we see that:
    \begin{equation}
      \frac{r_\star^{3/2}}{f_0^2 b_0} \gg 1 \,
    \end{equation}
    or:
    \begin{equation}
    \label{Cond2}
        n \gg \left( \frac{N_6}{F_0} \right)^{1/3} \, .
    \end{equation}
    The same condition ensures that the background curvature measured at $r=r_\star$ will be small at string units.
    \footnote{The Ricci scalar of the metric (\ref{TheBackground}) behaves like $\mathcal{R}  \sim h_6^{-3/2} \Box h_6$.}
   \item We have to check that the string coupling $e^\phi$ is small at $r=r_\star$:
   \begin{equation}
      f_0 r_\star^{-3/16} \gg 1 \, .
   \end{equation}
   This gives the third and the final restriction:
    \begin{equation}
    \label{Cond3}
        n \ll \left( \frac{N_6^5}{F_0} \right)^{1/3} \, .
    \end{equation}
\end{itemize}
Evidently we can always satisfy all four conditions (\ref{Cond4}), (\ref{Cond1}), (\ref{Cond2}) and (\ref{Cond3}) by adopting proper
values of $n$, $F_0$ and $N_6$.

At a first glance, the condition $N_6/F_0 \gg 1$ which is necessary for the validity of our calculation, appears to be opposite to the condition for the metastable state in the probe approximation of \cite{Kachru:2002gs}. Indeed, if one considers the polarization of $N_6$ anti-D3 branes into one NS5 brane, a metastable solution only exists when the number of anti-D3 branes divided by the $F_3$ flux on the conifold, $N_6 / M $, is less than $8\%$. However, if one considers the polarization of the anti-D3 branes into {\em multiple} NS5 branes, one can trivially extend the KPV calculation and show that there will exist metastable vacua for an arbitrarily-large number of anti-D3 branes, as long as  $N_6 / M $ divided by the number of these NS5 branes does not exceed $8\%$. Hence, our calculation rules out brane polarization even in the regime where a ``generalized KPV'' probe calculation would find it.

Although one may naively think that the no-polarization result we found may not be valid once we relax these conditions, it is very easy to see that this will not be so. In fact, these conditions are those that take us to the regime where polarization is {\em most} likely and relaxing them generically makes polarization less favored.
Indeed, the potential we computed has all the right features to allow for the branes to polarize, except that its coefficients (\ref{coefficients}) are not the good ones.

For example, as we explained above, to make polarization easier,  the energy of the probe must be dominated by the branes dissolved in it and the condition (\ref{Cond-DBI}) guarantees that. The condition that the probe has less D6 charge than the number of D6 branes sourcing the background (\ref{Cond4}) is necessary to do the probe calculation {\em \`a la Polchinski-Strassler}, but, as explained there, one can treat the full brane polarization problem by considering multiple shells with $n \ll N_6$ and treating each shell as a probe in the background of of the others. This ``mean-field'' calculation agrees perfectly with the fully-backreacted supergravity solution whenever the latter has been done. Hence, the fact that branes do not polarize for $n \ll N_6$ makes it highly implausible they will polarize otherwise.
Similarly, the condition that the radius of the D8 branes is larger than the string scale is only necessary for the validity of the DBI calculation; when this radius is smaller than the string scale one can still describe the polarization potential using the non-Abelian Born-Infeld action of the anti-D6 branes, and the result is the same \cite{Myers:1999ps}.

\subsection*{Acknowledgements}
We thank Anatoly Dymarsky, Mariana Gra\~na, Stefano Massai, Liam McAllister, Paul McGuirk and Gonzalo Torroba for useful and oftentimes challenging discussions. The work of I.B. and S.K. is supported in part by the ANR grant 08-JCJC-0001-0, and by the ERC Starting Independent Researcher Grant 240210 - String-QCD-BH. The work of D.J. and M.Z. is supported by the German Research Foundation (DFG) within
%the Emmy Noether Program (Grant number ZA 279/1-2) and
the Cluster of Excellence ``QUEST''. The work of T.V.R. is supported
by the ERC Starting Independent Researcher Grant 259133
-ObservableString. The work of T.W. is supported by a Research Fellowship (Grant
number WR 166/1-1) of the German Research Foundation (DFG), by the
Alfred P. Sloan Foundation and by the NSF under grant PHY-0757868. D.J., S.K. and T.V.R. would also like to thank the organizers of the Uppsala workshop on ``Brane Backreaction, Fluxes and Metastable Vacua in String Theory'' for providing an environment for stimulating discussions on this subject.

\appendix

\section{The compact solution}

In this Appendix we repeat our calculations to the compact solution of \cite{Blaback:2011pn}, which looks exactly like (\ref{TheBackground}) with $\mathbb{R}^{1,6}$ and $\mathbb{R}^{3}$ replaced by $AdS_{7}$ and $S^{3}$ respectively.

The relations that determine $a_1$, $b_1$ and $f_1$ are now:
\begin{eqnarray}
\label{a1b1f1COMPACT}
  \frac{a_1}{a_0} &=&  \frac{27}{16} \frac{a_0^2}{b_0} + \frac{1}{16} \frac{\lambda_1 + 1}{\lambda_0} - \frac{3}{64} \frac{(F_0 \lambda_0)^2}{b_0 f_0^{10}}
  \nonumber \\
  \frac{b_1}{b_0} &=& \frac{35}{8} \frac{a_0^2}{b_0} - \frac{7}{8} \frac{\lambda_1 + 1}{\lambda_0} + \frac{31}{96} \frac{(F_0 \lambda_0)^2}{b_0 f_0^{10}}
  \nonumber \\
  \frac{f_1}{f_0} &=& - \frac{63}{16} \frac{a_0^2}{b_0} + \frac{3}{16} \frac{\lambda_1 + 1}{\lambda_0} + \frac{7}{64} \frac{(F_0 \lambda_0)^2}{b_0 f_0^{10}}
\end{eqnarray}
As one can see, the non-compact relations (\ref{a1b1f1}) differ from these only by the first terms on the right hand side.
This is consistent with the fact that $a_0$ and $b_0$ parameterize the radius of $AdS_{7}$ and $S^{3}$, and so, at least formally,
the infinite radius limit should yield (\ref{a1b1f1}).

Next, (\ref{lambda2a2f2}) remains unmodified and with (\ref{a1b1f1COMPACT}) we find a different result for $c_2$:
\begin{eqnarray}
c_2  &=& - \frac{7}{a_0^5 b_0 f_0^3} + \frac{1}{12} \cdot \frac{(F_0 \lambda_0)^2}{a_0^7 b_0 f_0^{13}} \, ,
  \label{newc2} \\
c_3 &=& \frac{1}{3} \cdot \frac{F_0 \lambda_0}{a_0^7 b_0^{3/2} f_0^{10}} \, ,
   \\
c_4 &=&  \frac{1}{2} \cdot \frac{1}{a_0^7 b_0^{2} f_0^{7}} \, .
\end{eqnarray}
Clearly the extra term in $c_2 $ makes polarization more likely than in the non-compact solution. The strength of this term depends on the curvature of the $AdS_7$ and the 3-sphere and is not a free parameter, but rather fixed by the D6 charge, $g_s N_6$. Hence, the presence of polarization can only be determined by knowing the full compact solution.

Note that in realistic scenarios where a KS throat is embedded in a compact flux background, the length scale of the latter is not related to the number of anti-branes, and hence the first term in (\ref{newc2}) will be parametrically smaller than the other terms, and will not help in restoring brane polarization.

\bibliography{draft}

\bibliographystyle{utphysmodb}

\end{document}